\documentclass[showpacs,amsmath,amssymb,11pt]{revtex4} 
\usepackage{graphicx}% Include figure files 
\usepackage{dcolumn}% Align table columns on decimal point 
\usepackage{bm}% bold math 
\usepackage{amsmath} 

\begin{document} 
 
%\draft 
 
\title{\bf Enhancement of  nuclear Schiff moments and time 
 reversal violation in atoms due to combination of 
soft nuclear octupole and quadrupole vibrations} 
\author{A. F. Lisetskiy$^{1}$, V.V. Flambaum$^{2}$, and V.G. Zelevinsky$^{1}$} 
\address{$^{1}$National Superconducting Cyclotron Laboratory and\\ 
Department of Physics and Astronomy, Michigan State University, 
East Lansing, MI 48824-1321, USA\\ 
$^{2}$School of Physics, University of New South Wales, Sydney 
2052, Australia \\ 
} 
\date{\today}

%\tightenlines 
%***************************************************************** 
\begin{abstract} 
Nuclear forces violating parity and time reversal invariance 
(${\cal P},{\cal T}$-odd) produce ${\cal P},{\cal T}$-odd nuclear 
moments, for example, the nuclear Schiff moment. In turn, this 
moment can induce the electric dipole moment (EDM) in the atom. 
The contribution to the Schiff moment from the soft collective 
quadrupole and octupole vibrations in spherical nuclei is 
calculated in the framework of the quasiparticle random phase 
approximation with separable quadrupole and octupole forces. The 
values of nuclear Schiff moments predicted  for odd $^{217-221}$Ra 
and $^{217-221}$Ra isotopes indicate a possibility of enhancement 
by a factor of 50 or more as compared to the experimentally 
studied spherical nuclei $^{199}$Hg and $^{129}$Xe. Since the EDM 
in very heavy atoms, such as Ra, Rn, and Fr, has an additional 
enhancement rapidly increasing with nuclear charge $Z$, the EDM 
enhancement can exceed two orders of magnitude. We discuss the 
nuclear structure effects causing an enhancement of the Schiff 
moment. 
\end{abstract} 
 
%************************************************************************* 
 
\maketitle 
 
\section{Introduction} 
 
The search for interactions violating time reversal (${\cal T}$-) 
invariance is an important part of studies of fundamental 
symmetries in nature. The manifestations of ${\cal CP}$-violation 
(and therefore, through the ${\cal CPT}$-theorem, of ${\cal 
T}$-invariance) in systems of neutral $K$- and $B$-mesons 
\cite{fleischer02} set limits on physical effects beyond the 
standard model. However, the main hopes for the extraction of 
nucleon-nucleon and quark-quark interactions violating fundamental 
symmetries emerge from the experiments with atoms and atomic 
nuclei, see the recent review \cite{ginges04} and references 
therein. For example, the best limits on ${\cal P},{\cal T}$-odd 
forces have been obtained from the measurements of the atomic 
electric dipole moment (EDM) in the $^{199}$Hg \cite{romalis01} 
and $^{129}$Xe \cite{jacobs95} nuclei. As we know from past 
experience with ${\cal P}$-odd forces, see the review article 
\cite{flamgrib95}, there are powerful many-body mechanisms in 
heavy atoms and nuclei which allow one to expect a significant 
amplification of effects generated on the level of elementary 
interactions. There are also suggestions for using possible 
molecular and solid state enhancement mechanisms 
\cite{hudson02,lamoreaux02,mukhamedjanov05}. 
 
The theoretical calculation of atomic EDM proceeds through the 
{\sl nuclear Schiff moment} ${\bf S}$ since the nuclear EDM is 
shielded by atomic electrons \cite{FKS84}. The Schiff moment 
produces the ${\cal PT}$-odd electrostatic potential that, in 
turn, induces the atomic EDM. The expectation value of the vector 
operator ${\bf S}$ in a stationary nuclear state characterized by 
certain quantum numbers of angular momentum, $JM$, is possible 
only for $J\neq 0$ owing to the requirements of rotational 
invariance. Since all even-even nuclei have zero ground state 
spin, we need to consider an odd-$A$ nucleus. Furthermore, the 
non-zero expectation value of a polar vector ${\bf S}$ requires 
parity non-conservation in a nucleus; in addition, being 
proportional to the ${\cal T}$-odd pseudoscalar $\langle ({\bf 
S}\cdot{\bf J})\rangle$, this expectation value reveals the 
violation of ${\cal T}$-invariance. 
 
A reliable evaluation of the nuclear Schiff moment should include 
the estimates of renormalization effects due to ``normal" strong 
interactions inside the nucleus. The core polarization by the odd 
nucleon is important, especially in the case of the odd neutron, 
as $^{199}$Hg and $^{129}$Xe. Calculations 
\cite{FV,DS03,DS04} show that the resulting configuration 
mixing, depending on details of the method, may change the result 
of the independent particle model by a factor of about 2. A 
possibility of using accidental proximity of nuclear levels with 
the same spin and opposite parity was pointed out in Refs. 
\cite{FKS84,haxton83}. In such approaches, possible {\sl coherent} 
enhancement mechanisms are usually not considered. 
 
The statistical many-body enhancement of parity non-conservation 
in the region of the high level density of neutron resonances was 
predicted theoretically (see e.g. reviews \cite{SF82,flamgrib95} 
and references therein). The existence of such 
enhancement is now well documented experimentally 
\cite{mitchell}. The simultaneous violation of parity and time 
reversal invariance can be enhanced by the firmly established 
${\cal P}$-violation due to ${\cal P}$-odd ${\cal T}$-even weak 
interactions. The idea of a possible role of {\sl static octupole 
deformation} \cite{SF,SF82,FZ95,SA} exploited the parity doublets which 
appear in the presence of pear-shaped intrinsic deformation of the 
mean field. The doublet partners have similar structure and 
relatively close energies so that they can be more effectively 
mixed by ${\cal P}$-odd forces. The Schiff moment in the 
body-fixed frame is enhanced being in fact proportional to the 
collective octupole moment. The microscopic calculations 
\cite{AFS96,SAF97} predict a resulting Schiff moment by two-three 
orders of magnitude greater than in spherical nuclei; this enahncement 
 was confirmed  in Refs. \cite{engel03,DE}. The uncertainties 
related to the specific assumptions on ${\cal PT}$-odd forces and 
different approximations for nuclear structure are on the order of 
a factor 2 for the resulting Schiff moment. 
 
It was suggested in Ref. \cite{hayes00} that {\sl soft octupole 
vibrations} observed in some regions of the nuclear chart more 
frequently than static octupole deformation may produce a similar 
enhancement of the Schiff moment. This would make heavy atoms 
containing nuclei with large collective Schiff moments attractive 
for future experiments in search for ${\cal P,T}$-violation; 
experiments of this type are currently under progress or in 
preparation in several laboratories. Recently we performed 
\cite{FZ03} the estimate of the Schiff moment generated in nuclei 
with the soft octupole mode and showed that the result is nearly 
the same as in the case of the static octupole deformation. 
 
A related idea is explored in the present paper. It is known that 
some nuclei are soft with respect to {\sl both quadrupole and 
octupole} modes, see for example recent predictions for 
radioactive nuclei along the $N=Z$ line \cite{kaneko02}. The light 
isotopes of Rn and Ra are spherical but with a soft quadrupole 
mode and therefore large amplitude of quadrupole vibrations. The 
spectra of these nuclei display long quasivibrational bands 
\cite{jarmstrong99} based on the ground state and on the octupole 
phonon, with positive and negative parity, respectively. These 
bands are connected via low-energy electric dipole transitions. 
This situation seems to be favorable for the enhancement of ${\cal 
P},{\cal T}$-odd effects. 
 
Below we show that the  enhancement indeed exists in spherical 
nuclei which have both collective quadrupole and octupole modes. 
The main mixing occurs between the levels of the same spin and 
opposite parity that carry a significant admixture of the $2^{+}$ 
or/and $3^{-}$ phonons to the odd nucleon. In the odd-neutron 
nuclei the Schiff moment is originated by the proton contribution 
to the collective phonon. A number of such nuclei have an 
appropriate opposite parity level with the same angular momentum 
close to ground state. In this case the enhancement factor for the 
nuclear Schiff moment and atomic EDM may exceed 50. At the next 
stage the task of theory should be to combine all effects 
generated by ${\cal PT}$-violating interaction, including core 
polarization and weak interaction admixtures to the phonon 
structure. 
 
\section{Collective Schiff moment in spherical nuclei: 
A simple estimate} 
 
Consider a nucleus with two close levels of the same spin $J$ and 
opposite parity, ground state $|{\rm g.s.}\rangle$ and excited 
state $|x\rangle$. The energies of these states are $E_{{\rm 
g.s.}}$ and $E_{x}$, respectively. Let $W$ be a ${\cal P},{\cal 
T}$-odd interaction mixing the unperturbed states. Assuming that 
the mixing matrix elements of ${\bf S}$ and $W$ are real, we can 
write down the Schiff moment emerging in the actual mixed ground 
state as 
\begin{equation} 
{\bf S}= 2\frac{\langle {\rm g.s.}|W|x\rangle \langle x|{\bf 
S}|{\rm g.s.}\rangle}{E_{{\rm g.s.}}-E_{x}}. 
 \label{2.1} 
\end{equation} 
However, as it was explained in Ref. \cite{FKS84}, in the case of 
mixing of close {\sl single-particle} states one should not expect 
a large enhancement. For example, in a simple approximate model, 
where the strong nuclear potential is proportional to nuclear 
density and the spin-orbit interaction is neglected, the matrix 
element $\langle {\rm g.s.}|W|x\rangle$ contains the 
single-particle momentum operator. This matrix element is 
proportional to $(E_{{\rm g.s.}}-E_{x})$, so that the small energy 
denominator cancels out. As mentioned above, the collective Schiff 
moments in nuclei with static octupole deformations may be by 2-3 
orders of magnitude stronger than single-particle moments in 
spherical nuclei. 
 
The mechanism generating the {\sl collective} Schiff moment in 
spherical nuclei might be the following \cite{FZ03} (we illustrate 
the idea by an example). Let an odd-$A$ nucleus have the 
unperturbed ground state of spin $J$ built as a zero spin core 
plus an unpaired nucleon in the spherical mean field orbit with 
angular momentum $j=J$. The interaction between the odd particle 
and vibrations of the core causes an admixture of a quadrupole 
phonon to the ground state if the nuclear spin $J> 1/2$: 
\begin{equation} 
|{\rm g.s.}\rangle=a_{0}|j=J\rangle+a_{2}|[j,2^{+}]_J\rangle. 
                                                  \label{2.2} 
\end{equation} 
An opposite parity state can be formed by adding an octupole 
phonon to the ground state particle, 
\begin{equation} 
|x>=|[j,3^{-}]_J\rangle.                              \label{2.3} 
\end{equation} 
 
To find the ${\cal P},{\cal T}$-odd Schiff moment (\ref{2.1}) we 
need to know the matrix elements of the ${\cal P},{\cal T}$-odd 
nucleon-nucleon interaction. To the first order in the 
non-relativistic nucleon velocity $p/m$, the ${\cal P},{\cal 
T}$-odd interaction can be presented as \cite{FKS84} 
\begin{equation} 
W_{ab}=\frac{G}{\sqrt{2}}\frac{1}{2m}\Bigl( (\eta_{ab} 
\mbox{\boldmath$\sigma$}_{a}-\eta_{ba}\mbox{\boldmath$\sigma$}_{b})\cdot 
\mbox{\boldmath$\nabla$}_{a}\delta({\bf r}_{a}-{\bf r}_{b})+ 
\eta'_{ab}\left[ \mbox{\boldmath$\sigma$}_{a}\times 
\mbox{\boldmath$\sigma$}_{b}\right] \cdot \left\{({\bf p}_{a}-{\bf 
p}_{b}),\delta({\bf r}_{a}-{\bf r}_{b}) \right\}\Bigr), 
                                                   \label{2.4} 
\end{equation} 
where $\{\ ,\ \}$ is an anticommutator, $G$ is the Fermi constant 
of the weak interaction, $m$ is the nucleon mass, and 
\mbox{\boldmath$\sigma$}$_{a,b}$, ${\bf r}_{a,b}$, and ${\bf 
p}_{a,b}$ are the spins, coordinates, and momenta, respectively, 
of the interacting nucleons $a$ and $b$. The dimensionless 
constants $\eta _{ab}$ and $\eta '_{ab}$ characterize the strength 
of the ${\cal P},{\cal T}$-odd nuclear forces; in fact, 
experiments on measurement of the EDMs are aimed at extracting the 
values of these constants. 
 
A single-particle matrix element of $W$ has been estimated in Ref. 
\cite{SAF97}. Based on this experience, we present the matrix 
elements of $W$ between the states (\ref{2.2}) and (\ref{2.3}) 
dressed by the phonons as single-particle matrix elements times 
numerical factors $K_W$. 
\begin{equation} 
\langle {\rm g.s.}|W|x\rangle=\eta \frac{G}{2 \pi \sqrt{2} m r_0^4 
A^{1/3}} 
 K_W \approx \frac{\eta}{A^{1/3}} K_W \; {\rm eV};  \label{2.5} 
\end{equation} 
here $r_0 \approx 1.2$ fm is an internucleon distance, and $A$ is 
the mass number of the nucleus. By definition, $K_W \approx 1$ for 
single-particle matrix elements. Numerical calculations show that 
for the phonon-dressed close opposite parity states $K_W \approx 
0.3$. 
 
Similarly, we can present the matrix elements of the Schiff moment 
operator $S$ between the phonon-dressed states as single-particle 
matrix elements times numerical factors $K_S$, 
\begin{equation} 
\langle{\rm g.s.}|S_z|x\rangle=K_S\,e\,\cdot{\rm fm}^3.\label{2.6} 
\end{equation} 
Realistically, single-particle matrix elements $K_S$ are between 
0.3 and 2 when defined for the maximum projection $J_z$ of the 
angular momentum $J$. Numerical calculations has shown that for 
the phonon-dressed close opposite parity states $K_S \approx 1.0$. 
Although there is an enhancement of the Schiff moment matrix 
element between the quadrupole and octupole phonon states 
(enhancement factor $\sim 2$), that has a collective origin, the 
fragmentation of quasiparticle-phonon components and angular 
momentum recoupling reduce this factor typically to $K_S \approx 
1.0$. 
 
In the case of {\sl static} deformation, in the ``frozen" 
body-fixed frame the intrinsic collective Schiff moment $S_{\rm 
intr}$ of the deformed nucleus can exist without any ${\cal 
P},{\cal T}$-violation \cite{AFS96,SAF97} 
\begin{equation} 
S_{\rm intr} \approx \frac{9}{20\pi \sqrt{35}}\,eZR^{3}\beta 
_{2}\beta_{3}=\frac{3}{5\sqrt{35}}O_{\rm intr}\beta_{2} \ , 
                                         \label{2.7} 
\end{equation} 
where  $\beta_{2}$ is the static quadrupole and $\beta_{3}$ is the 
static octupole deformation parameters, $O_{\rm intr}$ is the 
static octupole moment. Of course, in the space-fixed laboratory 
frame, the nucleus has definite angular momentum rather than fixed 
orientation, and this makes the expectation value of the Schiff 
moment to vanish in the case of no ${\cal PT}$-violation. 
 
The relation (\ref{2.7}) should hold \cite{FZ03} for the {\sl 
dynamic} quadrupole and octupole deformations in systems with 
spherical equilibrium shape. Using eqs. (\ref{2.1},\ref{2.5}), and 
(\ref{2.6}) we can find the ground state Schiff moment: 
\begin{equation} 
S=\frac{\eta}{E_{{\rm g.s.}}-E_{x}}\, \frac{G}{\pi \sqrt{2} m 
r_0^4 A^{1/3}} K_W K_S\,e\,\cdot{\rm fm}^3 \approx 10^{-5} \eta\, 
\frac{100 \;{\rm keV}} {E_{{\rm g.s.}}-E_{x}} \frac{2 K_S 
K_W}{A^{1/3}}\,e\,{\rm fm}^3.              \label{2.8} 
\end{equation} 
For both single-particle matrix elements and collective matrix 
elements between the phonon states $K_S K_W \approx 0.3$. Therefore, 
in the case of close levels of opposite parity ( $E_{{\rm 
g.s.}}-E_{x} \sim$ 100 keV), the Schiff moment exceeds that of 
$^{199}$Hg by two orders of magnitude: 
\begin{equation} 
S\approx 100\cdot 10^{-8}\eta\,\frac{100\;{\rm keV}}{E_{{\rm 
g.s.}} -E_{x}}\,e\,{\rm fm}^{3} \sim 70\; S(^{199}{\rm Hg}). 
\label{2.9} 
\end{equation} 
Here we used the Schiff moment value from Ref. \cite{FKS86}, 
$S(^{199}$Hg$)=-1.4\cdot 10^{-8} \eta \,e\,\cdot{\rm fm}^3$. 
 
Below we give the details of the calculations in the framework of 
QRPA for quadrupole and octupole phonons. 
 
\section{Description of calculations} 
 
\subsection{QRPA phonons and particle-phonon interaction} 
 
We start with the quadrupole ($J^\pi =2^+$) and octupole ($J^\pi 
=3^-$) phonon states in even-even nuclei based on a conventional 
model Hamiltonian, 
\begin{equation} 
H=H_{\rm s-p} + H_{\rm pair} + H_{\rm phon}.    \label{3.1.1} 
\end{equation} 
The first term, $H_{\rm s-p}$, describes the Woods-Saxon 
potential. The following parameterization was used: 
$$V_c^{p(n)}=-49.6\left(1+(-)0.86\,{N-Z \over A} \right) {\rm MeV}$$ 
and $$V_{ls}=18.8\left({A \over A-1}\right)^2 {\rm MeV}$$ are the 
strength parameters for the central and spin-orbital potentials, 
respectively; $R=1.3\,A^{1/3}$ fm and $a=0.7$ fm are the radius 
and diffuseness parameters.   The second term, $H_{\rm pair}$, is 
pairing interaction with the pairing strength 
$$G_p ={17.9 +0.176\,(N-Z) \over A}$$ for protons and 
$$G_n = {18.95 -0.078\,(N-Z) \over A}$$ 
for neutrons. The BCS formalism was used that yields a 
quasiparticle basis with corresponding Bogoliubov transformation 
coefficients $u_j$ and $v_j$.  The last term, $H_{\rm phon}$, 
presents RPA phonons obtained with a separable multipole-multipole 
interaction, 
\begin{equation} 
H_{\rm phon}=\sum_{\lambda \mu}\omega_\lambda Q^{\dagger}_{\lambda 
\mu} Q_{\lambda \mu}.                               \label{3.1.2} 
\end{equation} 
The building blocks of the model are the two-quasiparticle RPA 
phonons, 
\begin{equation} 
Q^\dagger_{\lambda,\mu}=\frac{1}{2} \sum_{j_1,j_2}\left[ 
A_{j_1,j_2}^{\lambda} 
[\alpha^\dagger_{j_1}\alpha^\dagger_{j_2}]_{\lambda,\mu} 
-(-)^{\lambda-\mu} B_{j_1,j_2}^{\lambda} 
[\alpha_{j_1}\alpha_{j_2}]_{\lambda,\mu} \right],   \label{3.1.3} 
\end{equation} 
where $A_{j_1,j_2}$ and $B_{j_1,j_2}$ are the forward and backward 
phonon amplitudes, 
\begin{equation} 
A_{j_1,j_2}^\lambda = \frac{1}{\sqrt{2Z(\lambda)}}\, 
\frac{f^\lambda(j_1j_2)u^{(+)}_{j_1,j_2}}{ 
\varepsilon(j_1,j_2)-\omega_{\lambda}}; \quad B_{j_1,j_2}^\lambda 
= \frac{1}{\sqrt{2Z(\lambda)}}\, 
\frac{f^\lambda(j_1j_2)u^{(+)}_{j_1,j_2}}{ 
\varepsilon(j_1,j_2)+\omega_{\lambda}},         \label{3.1.4} 
\end{equation} 
where $Z(\lambda)$ is a normalization factor, $f^\lambda(j_1j_2)= 
<j_1||r^\lambda Y_{\lambda ,\mu }||j_2>$, and coherence factors of 
the Bogoliubov canonical transformation are $u^{(\pm)}_{1,2}= 
u_1v_2 \pm v_1u_2$ and $v^{(\pm)}_{1,2}=u_1u_2 \pm v_1v_2$. 
 
The phonon frequencies $\omega_{\lambda}$ are the roots of the 
characteristic RPA equations for each multipolarity $\lambda$, 
\begin{equation} 
X(\lambda) \equiv \frac{1}{2\lambda+1} \sum_{j_1,j_2} 
\frac{[f^\lambda(j_1j_2)u^{(+)}_{j_1,j_2}]^2 
\varepsilon(j_1,j_2)}{ \varepsilon^2(j_1,j_2)-\omega_{\lambda}^2} 
=\frac{1}{\chi(\lambda)},                       \label{3.1.5} 
\end{equation} 
where $\chi(\lambda)$ is the strength parameter and 
$\varepsilon(j_1,j_2)=\varepsilon(j_1)+ \varepsilon(j_2)$ is the 
unperturbed two-quasiparticle energy. Solving these equations one 
obtains the energies of the phonons and internal structure of the 
phonon operator (\ref{3.1.3}) hidden in the amplitudes of 
different two-quasiparticle components. The values of the strength 
parameters, $\chi(2)=0.0187$ and $\chi(3)=0.0019$, were chosen to 
reproduce the excitation energies of the $J^\pi=2^+_1$ and the 
$J^\pi=3^-_1$ states in $^{212-218}$Ra and Rn isotopes. 
 
Since we are interested in the evaluation of the Schiff moment of 
the odd-neutron nuclei, the second step of calculations reduces to 
the solution of the secular equation for the odd-$A$ nucleus with 
the even-even core excitations described above. If we neglect the 
quasiparticle+two-phonon components in the wave function of an 
excited state of an odd-A nucleus, then the corresponding wave 
function has the following form: 
\begin{equation} 
\Psi_n(J^\pi)=C_{J}\Bigl(\alpha_{JM}^{\dagger} + 
\sum_{\lambda,j}D_{Jj}^{\lambda,n} [\alpha_{j}^{\dagger}\otimes 
Q^+(\lambda)]_{JM}\Bigr)\Psi_0.                 \label{3.1.6} 
\end{equation} 
 
The energy $E_n(J)$ of the $n^{{\rm th}}$ state with angular 
momentum $J$ in the odd-mass nucleus, the amplitudes of the 
quasiparticle-phonon components, 
\begin{equation} 
D_{Jj}^{\lambda,n}=\sqrt{\frac{2\lambda+1}{(2J+1)2Z(\lambda)}}\, 
\frac{f^\lambda(Jj)v^{(-)}_{J,j}}{ 
\varepsilon(j)+\omega_{\lambda}-E_n},            \label{3.1.7} 
\end{equation} 
and the amplitude of the single-quasiparticle component, 
\begin{equation} 
C_{J,n}^{-2}=1+\sum_{\lambda, j} (D_{Jj}^{\lambda,n})^2, 
                                                 \label{3.1.8} 
\end{equation} 
are obtained from the solution of the following secular equation 
\cite{Sol92}: 
\begin{equation} 
\varepsilon(J) - E_n =\sum_{\lambda,j} 
\left(D_{Jj}^{\lambda,n}\right)^2 \left[ 
\varepsilon(j)+\omega_{\lambda}-E_n\right].    \label{3.1.9} 
\end{equation} 
 
The quasiparticle-phonon structures of the wave functions for the 
odd-neutron radium and radon isotopes with A=217, 219 and 221 are 
shown in Table I. 
 
%****************************************** 
\subsection{Calculation of the Schiff Moment} 
%****************************************** 
 
Having determined the wave-functions one may proceed with the 
calculation of the Schiff moment according to the Eq. (\ref{2.1}), 
where for the ground state we will use notation $J_g^\pi$ and for 
the first excited state of opposite parity $J_g^{-\pi}$: 
\begin{equation} 
S(J_g^\pi)=2\, \frac{ \langle J_g^\pi | W | J_g^{-\pi}  \rangle 
\langle J_g^{-\pi} J_g | S_z | J_g^\pi J_g \rangle}{E (J_g^\pi) - E 
(J_g^{-\pi} )}, \label{3.2.1} 
\end{equation} 
where $W$ is the ${\cal P,T}$-violating nucleon-nucleon 
interaction given by the Eq. (\ref{2.4}) and $\langle J_g^{-\pi} 
J_g | S | J_g^\pi J_g\rangle$ is the matrix element of the 
Schiff operator ${\bf S}$, 
\begin{equation} 
S_\mu=\frac{1}{10}\sqrt{4\pi \over 3}\, 
\sum_iY_{1\mu}(\theta_i,\varphi_i)e_{i}\left[ r_i^{3} 
-\frac{5}{3Z}\, r^{2}_{\rm ch}r_i \right], \label{3.2.2} 
\end{equation} 
with the maximum projection $M=J_g$ of angular momentum and  
$r^{2}_{\rm ch}$ is the mean square charge radius. 
 
Since only the proton components of the phonons contribute to the 
matrix element of the Schiff moment, this matrix element for the 
case of the odd neutron can be written as 
\begin{equation} 
\langle J^\pm_g J_g| S_z | J^\mp_g J_g \rangle = 
\sqrt{J_g (2J_g+1) \over (J_g+1)} 
\sum_{j,\lambda,\lambda'} 
 (-1)^{j+\lambda'+J_g+1} 
D^\lambda_{J_g^\pm,j}D^{\lambda'}_{J_g^\mp,j} \left\{ 
\begin{array}{rrr} 
  \lambda & \lambda' &  1    \\ 
  J_g & J_g & j \\ 
  \end{array} 
\right\} \langle \lambda \| S \| \lambda' \rangle, \label{3.2.3} 
\end{equation} 
where the sum over $\lambda$ and $\lambda'$ is reduced to one term 
with $\lambda^\pi=2^+$ and $\lambda'^\pi=3^-$. Specifically, the 
matrix element of the Schiff operator between the quadrupole and 
octupole phonon states has the following form in terms of the RPA 
amplitudes (\ref{3.1.4}): 
\begin{equation} 
\langle 2^+| S | 3^- \rangle = 
\sqrt{35}\sum_{j_1,j_2,j_3}v^{(-)}_{j_1,j_2} \langle j_1 \|S \|j_2 
\rangle \left\{ 
\begin{array}{rrr} 
  2 & 3 &  1    \\ 
  j_1 & j_2 & j_3 \\ 
  \end{array} 
\right\} (A^{(2+)}_{j_2 j_3}A^{(3-)}_{j_3 j_1} + B^{(2+)}_{j_2 
j_3}B^{(3-)}_{j_3 j_1}).             \label{3.2.4} 
\end{equation} 
The results of the calculations are shown in Table I. 
%**************************************************************** 
\begin{table} 
\label{strx} \caption{Calculated structure of the wave functions 
and the  matrix element of the Schiff operator. The 
dominant quasiparticle and quasiparticle-phonon components for the 
ground state ($J^{\pi}$) and the first excited state ($J^{-\pi}$) 
of opposite parity are shown.} 
\begin{center} 
\begin{tabular}{c|c|cc|c|cc|c} 
\hline 
 Nucleus  & \multicolumn{6}{c|}{ Structure } & 
$\langle J^\pi J | S_z | J^{-\pi} J  \rangle$ \\ 
\hline 
          & $J^\pi$  & $q.p.$ & $q.p.\otimes \lambda^\pi $ & $J^{-\pi}$ 
& $q.p.$   & 
              $q.p.\otimes \lambda^\pi$ & (e fm$^3$)  \\ 
\hline $^{217}_{88}$Ra$_{129}$ & $9/2^+$ & (82$\%$) $g_{9\over 2}$ 
& (10$\%$) $[g_{9\over 2}\otimes 2^+]$ & $9/2^-$ & 
(0.01$\%$) $h_{9\over 2}$ & (0.01$\%$) $[f_{5\over 2}\otimes 2^+]$ & 0.47 \\ 
      &  &               & (5$\%$) $[j_{15\over 2}\otimes 3^-]$ &   & 
                        & (99$\%$) $[g_{9\over 2}\otimes 3^-]$ & \\ 
\hline $^{217}_{86}$Rn$_{131}$ &  $9/2^+$ & (77$\%$) $g_{9\over 
2}$ & (12$\%$) $[g_{9\over 2}\otimes 2^+]$ & $9/2^-$ & 
(0.02$\%$) $h_{9\over 2}$ & (0.01$\%$) $[f_{5\over 2}\otimes 2^+]$ & 0.69   \\ 
      &  &               & (6$\%$) $[j_{15\over 2}\otimes 3^-]$ &   & 
                        & (99$\%$) $[g_{9\over 2}\otimes 3^-]$ & \\ 
\hline $^{219}_{88}$Ra$_{131}$ &   $7/2^+$ & (5$\%$) $g_{7\over 
2}$ & (85$\%$) $[g_{9\over 2}\otimes 2^+]$ & $7/2^-$ & 
(0.4$\%$) $f_{7\over 2}$ & (0.3$\%$) $[p_{3\over 2}\otimes 2^+]$ & 1.80 \\ 
      &  &               & (0.04$\%$) $[h_{11\over 2}\otimes 3^-]$ & & 
                        & (99$\%$) $[g_{9\over 2}\otimes 3^-]$ & \\ 
\hline $^{219}_{86}$Rn$_{133}$ &   $5/2^+$ & (25$\%$) $d_{5\over 
2}$ & (64$\%$) $[g_{9\over 2}\otimes 2^+]$ &  $5/2^-$ & 
(1.2$\%$) $f_{5\over 2}$ & (0.24$\%$) $[p_{1\over 2}\otimes 2^+]$ & 1.37  \\ 
      &   &              & (0.6$\%$) $[h_{11\over 2}\otimes 3^-]$ & & 
                        & (98$\%$) $[g_{9\over 2}\otimes 3^-]$ &\\ 
\hline $^{221}_{88}$Ra$_{133}$ &   $5/2^+$ & (33$\%$) $d_{5\over 
2}$ & (57$\%$) $[g_{9\over 2}\otimes 2^+]$ & $5/2^-$ & 
(2.3$\%$) $f_{5\over 2}$ & (0.6$\%$) $[p_{1\over 2}\otimes 2^+]$ & 2.21 \\ 
      &  &               & (1.0$\%$) $[h_{11\over 2}\otimes 3^-]$ & & 
                        & (95$\%$) $[g_{9\over 2}\otimes 3^-]$ & \\ 
\hline $^{221}_{86}$Rn$_{135}$ &   $7/2^+$ & (18$\%$) $g_{7\over 
2}$ & (22$\%$) $[g_{9\over 2}\otimes 2^+]$ &  $5/2^-$ & 
(0.7$\%$) $f_{7\over 2}$ & (0.2$\%$) $[p_{3\over 2}\otimes 2^+]$ & 1.03  \\ 
      &   &              & (0.2$\%$) $[j_{13\over 2}\otimes 3^-]$ & & 
                        & (99$\%$) $[g_{9\over 2}\otimes 3^-]$ &\\ 
\hline $^{211}_{86}$Rn$_{125}$ &   $1/2^-$ & (91$\%$) $p_{1\over 
2}$ & (5$\%$) $[f_{5\over 2}\otimes 2^+]$ &  $1/2^+$ & 
(0.2$\%$) $s_{1\over 2}$ & (0.01$\%$) $[d_{5\over 2}\otimes 2^+]$ & 0.16  \\ 
      &   &              & (0.4$\%$) $[g_{7\over 2}\otimes 3^-]$ & & 
                        & (99.7$\%$) $[f_{5\over 2}\otimes 3^-]$ &\\ 
\hline 
\end{tabular} 
\end{center} 
\end{table} 
%******************************************************************** 
 
The matrix element of the first term of the interaction in Eq. 
(\ref{2.4}) gives a dominant contribution that may be represented 
as 
\begin{equation} 
\langle ab;J^\pm M | W| cd;J^\mp M \rangle =\frac{G}{\sqrt{2}}{1 
\over 2m_p}\eta\,\langle ab;J^\pm M | W_1| cd;J^\mp M \rangle, 
                                                \label{3.2.5} 
\end{equation} 
 where 
\begin{equation} 
\langle ab;J^\pm M | W_1 | cd;J^\mp M \rangle = 
 F_1(abcd)\cdot K_1(ab;cd;J) +  F_0(abcd)\cdot K_0(ab;cd;J) 
                                            \label{3.2.6} 
\end{equation} 
for the interaction in the proton-neutron channel, while for the 
neutron-neutron channel it is given by 
\begin{equation} 
\langle ab;J^\pm M | W_1 | cd;J^\mp M \rangle = 
\frac{K_0(ab;cd;J)}{\sqrt{(1+\delta_{ab})(1+\delta_{cd})}} \left[ 
F_0(ab;cd)-(-1)^{J-j_c-j_d}F_0(ab;dc) \right]. \label{3.2.7} 
\end{equation} 
The coefficients $F_0$ and $F_1$ in Eqs. (\ref{3.2.6}) and 
(\ref{3.2.7}) represent radial parts of the matrix elements, 
\begin{equation} 
F_1(ab;cd)=I_r(abcd)(s_a+s_b+s_c+s_d+2),       \label{3.2.8} 
\end{equation} 
\begin{equation} 
F_0(ab;cd)=I_r(abcd)(s_a-s_b+s_c-s_d)+I_{ac}(bd)-I_{bd}(ac), 
                                              \label{3.2.9} 
\end{equation} 
where 
\begin{equation} 
I_r(abcd)=\int R_aR_bR_cR_drdr, \quad I_{ac}(bd)=\int 
[R_aR_c]'R_bR_dr^2dr,                         \label{3.2.10} 
\end{equation} 
and $s_a=(l_a-j_a)(2j_a+1)$. The angular $J$-dependent part of the 
two-body matrix element is given by the coefficients 
\begin{equation} 
 K_0(ab;cd;J)  = (-1)^{j_b-j_d+l_b+l_d}\Pi(abcd) 
\left( 
\begin{array}{rrr} 
  j_a & j_b &  J    \\ 
  {1 \over 2} & -{1 \over 2} & 0  \\ 
  \end{array} \right) 
\left( 
\begin{array}{rrr} 
  j_c & j_d &  J   \\ 
  {1 \over 2} & -{1 \over 2} & 0  \\ 
  \end{array} \right)                        \label{3.2.11} 
\end{equation} 
and 
\begin{equation} 
 K_1(ab;cd;J)  = \Pi(abcd) 
\left( 
\begin{array}{rrr} 
  j_a & j_b &  J    \\ 
  {1 \over 2} & {1 \over 2} & -1  \\ 
  \end{array} \right) 
\left( 
\begin{array}{rrr} 
  j_c & j_d &  J   \\ 
  {1 \over 2} & {1 \over 2} & -1  \\ 
  \end{array} \right).                           \label{3.2.12} 
\end{equation} 
where $\Pi(abcd)=\sqrt{(2j_a+1)(2j_b+1)(2j_c+1)(2j_d+1)}/8\pi$. 
 
The final expression for the interaction matrix element between 
the states $J_g^\pi$ and $J_g^{-\pi}$ may be represented as a sum 
of two contributions, 
\begin{equation} 
\langle J^\pi_g| W| J^{-\pi}_g \rangle = W_{q.p.} + W_{q.p.\otimes 
ph}.                                              \label{3.2.13} 
\end{equation} 
The first item originates from the interaction of the 
quasiparticle with the even-even core, 
\begin{equation} 
W_{q.p.} = C_{j_a=J^\pi_g}C_{j_{-a}=J^{-\pi}_g}w(a,-a) + 
\sum_{\lambda,b,-b}D^\lambda_{J^\pi_g,j_b}D^\lambda_{J^{-\pi}_g,j_{-b}} 
w(b,-b),                                       \label{3.2.14} 
\end{equation} 
where 
\begin{equation} 
w(a,-a)=\sum_{j_c,J} \frac{(2J+1)}{(2j_{a}+1)}W_{q.p.}(ac;-ac;J). 
                                                \label{3.2.15} 
\end{equation} 
Here the two-quasiparticle matrix element $W_{q.p}(ac;-ac;J_{ac})$ 
is a combination of the antisymmetrized two-particle interaction 
matrix elements weighted with the occupation factors of the 
single-particle states, 
\begin{equation} 
W_{q.p.}(ab;cd;J) = U_0(ab;cd) \langle j_aj_b;J | W | j_c j_d ;J 
\rangle^{as} +                              \label{3.2.16} 
\end{equation} 
\begin{equation*} 
\frac{ \left[ U_1(ab;cd) \langle j_aj_b^{-1};J | W| j_c j_d^{-1};J 
\rangle -(-1)^{j_c+j_d-J} U_2(ab;cd)\langle j_aj_b^{-1};J | W | 
j_d j_c^{-1};J \rangle \right] } 
{\sqrt{(1+\delta_{ab})(1+\delta_{cd})}}, 
\end{equation*} 
where the notations are used 
$U_0(ab;cd)=u_au_bu_cu_d+v_av_bv_cv_d$, 
$U_1(ab;cd)=u_av_bu_cv_d+v_au_bv_cu_d$, 
$U_2(ab;cd)=u_av_bv_cu_d+v_au_bu_cv_d$, and the two-body matrix 
elements in the particle-hole channel are related to the ones in 
the particle-particle channel through the Pandya relation, 
\begin{equation} 
\langle j_aj_b^{-1};J | W| j_c j_d^{-1};J \rangle = 
(-1)^{j_a+j_b+j_d+j_c+1}\times                \label{3.2.17} 
\end{equation} 
\begin{equation*} 
\sum_{J_1}(2J_1+1)\left\{ 
\begin{array}{rrr} 
  j_a & j_b &  J    \\ 
  j_c & j_d & J_1  \\ 
  \end{array} 
\right\}\langle j_dj_a;J_1 | W| j_b j_c ;J_1 \rangle. 
\end{equation*} 
 
The simplicity of the angular part of the interaction allows one 
to perform the summation over $J$ in Eq. (\ref{3.2.15}) 
analytically. The result for $w(a,-a)$ has the following form: 
\begin{equation} 
w(a,-a) = -{1 \over 4\pi } \sum_{j_c} (2j_c+1)\left[ 
I_{cc}(a,-a)+\frac{I_r(ac,-ac)(s_c+1)+I_{cc}(a,-a)}{\sqrt{1+ 
\delta_{ac}+\delta_{-a,c}}}\right]v^{(+)}_{a,-a}. \label{3.2.18} 
\end{equation} 
 
The second term in Eq. (\ref{3.2.14}) is due to the interaction of 
the odd quasiparticle with quadrupole and octupole phonons: 
\begin{equation} 
W_{q.p.\otimes ph} = \sum_{a,b,c,d,e,\lambda,\lambda'} 
 (-1)^{j_a+j_c+\lambda+\lambda'+1}\Bigl(A^{(\lambda)}_{j_b j_e} 
A^{(\lambda')}_{j_d j_e} + B^{(\lambda)}_{j_b j_e} 
B^{(\lambda')}_{j_d j_e}\Bigr)D^\lambda_{J_g^\pi,j_a} 
D^{\lambda'}_{J_g^{-\pi},j_c}  \times          \label{3.2.19} 
\end{equation} 
\begin{equation*} 
\left[ \delta_{a,c}\delta_{j_b,j_d}w(b,-b) + \sum_{J12} 
\Lambda(J_{12}J_{12}\lambda \lambda') \left\{ 
\begin{array}{rrr} 
  j_b & j_a &  J_{12}    \\ 
  J_g & j_e & \lambda  \\ 
  \end{array} 
\right\} \left\{ 
\begin{array}{rrr} 
  j_d & j_c &  J_{12}    \\ 
  J_g & j_e & \lambda'  \\ 
  \end{array} 
\right\}W_{q.p.}(ab,cd;J_{12}) \right], 
\end{equation*} 
where $\Lambda(J_{12}J_{12} \lambda 
\lambda')=(2J_{12}+1)\sqrt{(2\lambda+1) (2\lambda'+1)}$. The 
results for the interaction matrix element and final calculated 
values of the Schiff moment are shown in Table II. 
 
\begin{table} 
\label{resf} \caption{Calculated energy difference between the 
$J^\pi_g$ and closest $J^{-\pi}_g$ state of opposite parity 
($\Delta E_{+,-}^{\rm th}$), single-particle ($W_{q.p.}$) and 
collective ($W_{q.p.\otimes ph}$)  contributions to the total 
matrix element of the ${\cal PT}$-odd interaction, Eq. 
(\ref{3.2.12}), and corresponding value of the Schiff moment, Eq. 
(\ref{3.2.1}). The corresponding values of the atomic dipole 
moment d$_{\rm atomic}$ calculated using relations between Schiff 
and dipole moments \protect \cite{dzuba00} are listed in the last 
column. Previous measurements of   d$_{\rm atomic}$ were performed 
for Hg and Xe where $S \sim 1$ and  d$_{\rm atomic}\sim 1$ in the same units.} 
\begin{center} 
\begin{tabular}{c|c|c|cc|ccc} 
 Nucleus  & $J_g^\pi$ & $\Delta E_{+,-}^{\rm th}$  & $W_{q.p.}$ & $W_{q.p.\otimes ph}$ & $S(J_g^\pi)$ & d$_{\rm atomic}$ \\ 
\hline 
          &           &  (keV) &  ($\eta\cdot 10^{-2}$eV) & ($\eta\cdot 10^{-2}$eV) & 
($\eta\cdot 10^{-8}$ e$\cdot$ fm$^3$) & ($\eta\cdot 10^{-25}$ e$\cdot$ cm) \\ 
\hline 
$^{217}_{88}$Ra$_{129}$ & ${9 \over 2}^+$ &  1384 & -5.9$\cdot 10^{-2}$ & -2.3 & -1.58 & 13.4 \\ 
$^{217}_{86}$Rn$_{131}$ & ${9 \over 2}^+$ &  1296 &  7.9$\cdot 10^{-2}$ & -3.1 & -3.31 & -10.9\\ 
$^{219}_{88}$Ra$_{131}$ & ${7 \over 2}^+$ &  433  &  3.6$\cdot 10^{-2}$ & -8.1 & -70.1 & 595.9 \\ 
$^{219}_{86}$Rn$_{133}$ & ${5 \over 2}^+$ &  969  &  3.4$\cdot 10^{-2}$ & -6.5 & -18.5 &-61.1 \\ 
$^{221}_{88}$Ra$_{133}$ & ${5 \over 2}^+$ &  991  &  2.3$\cdot 10^{-2}$ & -11.5 & -51.7 &439.5 \\ 
$^{221}_{86}$Rn$_{135}$ & ${7 \over 2}^+$ &  567  &  7.9$\cdot 10^{-2}$ & -4.3 & -15.6 & -51.5 \\ 
$^{211}_{86}$Rn$_{125}$ & ${1 \over 2}^-$ &  2222  &  -2.6$\cdot 10^{-2}$ & -1.6 & -0.23 & -0.8 \\ 
\hline 
\end{tabular} 
\end{center} 
\end{table} 
 
\section{Discussion} 
 
We analyze here the results of calculations presented in a 
previous section. As seen from Eq. (19), there are three 
quantities determining the value of Schiff moment: an off-diagonal 
matrix element of the Schiff operator, 
$\langle J_g^{-\pi} M=J_g | S_z | 
J_g^\pi M=J_g \rangle$, a matrix element of  ${\cal PT}$-violating 
interaction, $\langle J_g^\pi | W | J_g^{-\pi}\rangle$, and 
excitation energy of the admixed state of opposite parity, $E 
(J_g^\pi) - E(J_g^{-\pi} )$. Let us discuss these quantities 
separately. 
 
The off-diagonal matrix elements of the Schiff operator, see Table 
1, are clearly correlated with the amplitudes of 
particle-plus-phonon components, Eq.~(21). In all cases under 
consideration, the neutron $g_{9/2}$ orbital has lowest energy and 
produces dominant particle-quadrupole and particle-octupole 
amplitudes. In the case when the ground state has spin 
$J^\pi=9/2^+$ (i.e., $^{217}$Ra and $^{217}$Rn), the single 
neutron quasiparticle  $g_{9/2}$ orbital dominates the ground 
state wave function. However, here there is only a contribution of 
small quasiparticle-phonon components leading to a small total 
matrix element of the Schiff operator. 
 
As soon as the ground state has a spin value different from the 
spin of the $g_{9/2}$ orbital, the particle-quadrupole phonon 
component becomes dominant for the ground state, and the matrix 
element grows. One may notice also that the Schiff matrix element 
increases with the mass number even if the 
quasiparticle-quadrupole- phonon component becomes smaller (e.g., 
$^{221}$Ra vs. $^{219}$Ra). This is caused by the enhancement of 
collectivity towards the middle of the shell, where, for example, 
the $^{222-224}$Ra isotopes have already well pronounced deformed 
structure. Thus, the matrix elements of the Schiff operator are 
sensitive to details of nuclear structure and under favorable 
conditions these matrix elements may be enhanced by a factor of 
two to four, as it was mentioned in Sec. II. 
 
The next important ingredient to the Schiff moment is a matrix 
element of ${\cal PT}$-violating interaction $\langle J_g^\pi | W 
| J_g^{-\pi}  \rangle$. We have split it into two parts, a single 
quasiparticle, $W_{q.p.}$, and collective, $W_{q.p.\otimes ph.},$ 
contributions shown in Table II. Since the amplitude of the 
single-quasiparticle component is quite small for the excited 
state of negative parity, see Table I, the $W_{q.p.}$ part is by 
two-three orders of magnitude smaller than the estimated 
single-particle value, $\eta A^{-1/3}$ eV ($\eta\cdot$0.16 eV for 
$A=221$, for example). Therefore the main contribution comes from 
collective quasiparticle-phonon components. The collective 
contribution to the interaction is correlated with the matrix 
element of the Schiff operator: stronger interaction corresponds 
to a larger off-diagonal matrix element of the Schiff operator. In 
most cases the interaction matrix element is suppressed compared 
to its single-particle estimate; similar observation has been 
reported \cite{engel03} in the case of Skyrme-Hartree-Fock 
calculations for deformed nuclei. However, there is a combined 
enhancement of the numerator in Eq.~(19) which may differ by a 
factor of 15 for neighboring isotopes, as, for example, $^{217}$Ra 
and $^{219}$Ra. 
 
Finally, there is a third factor, excitation energy of the 
opposite parity state, that enters the expression for the Schiff 
moment in the denominator. Coupling of the positive parity 
quasiparticle to the octupole phonon results in the gap $\Delta 
E_{+,-}$ that is at least by a factor of four smaller than the gap 
between single-particle partners with the same spin and opposite 
parity. If there is the same single-particle orbital (in our case 
$g_{9/2}$) coupled to quadrupole and octupole phonons, then the 
gap is closely correlated with the distance between the $2^+$ and 
$3^-$ states: the smaller it is the smaller is the value of 
$\Delta E_{+,-}$ and the larger is the enhancement factor for the 
Schiff moment. The $\Delta E_{+,-}$ gaps obtained in the framework 
of the QRPA are shown in Table II. The calculations of the Schiff 
moment with these energies yield the maximum value of 70 $\eta 
10^{-8}e\cdot\,$fm$^3$ for $^{219}$Ra and slightly smaller one, 52 
$\eta 10^{-8}e\cdot\,$fm$^3$, for $^{221}$Ra. Unfortunately, 
experimental data for the negative parity states for the nuclei 
discussed here are rather scanty. However, there are candidates 
for the $5/2^-$ state in $^{221}$Ra at 104 keV and 450 keV. If 
these states have needed nature, then the Schiff moment may be 
around 450 $\eta 10^{-8}e\cdot\,$fm$^3$  and 100 $\eta 
10^{-8}e\cdot\,$fm$^3$, respectively. A similar situation exists 
for $^{219}$Rn, where the $5/2^-$ candidate is 500 keV lower than 
the corresponding calculated $5/2^-$ state. Taking into account 
that we have used the QRPA in its simplest version, one may expect 
stronger enhancement revealed by more realistic calculations. 
 
Another interesting example is $^{211}$Rn (see Table I and II). 
The calculated Schiff moment in this case is even smaller than the 
estimated single-particle value. This is caused by the fact that 
the ground $J^\pi=1/2^-$ state has almost pure single-particle 
$p_{1/2}$ nature, while the first excited  $J^\pi=1/2^+$ state is 
calculated to be a pure quasiparticle-phonon $f_{5/2}\otimes 3^-$ 
state. Furthermore, the $3^-$ state is predicted to be high (1.7 
MeV) and almost non-collective. There is no chance to get here any 
enhancement of the Schiff moment along the lines of our 
calculations. The coupling of the quasiparticle to other low-lying 
negative parity states may change the result although keeping the 
value on the same order of magnitude. The main contribution in 
this case is due to the polarization of the proton charge density 
in the core by the ${\cal PT}$-odd field (\ref{2.4}) of the 
external neutron, the same mechanism that produces the Schiff 
moment of $^{199}$Hg \cite{FKS86}. Therefore, $S(^{211}$Rn$)\sim 
S(^{199}$Hg), and $d_{{\rm atomic}}(^{211}$Rn$)\sim d_{{\rm 
atomic}}(^{199}$Hg). 
 
\section{Conclusion} 
 
To conclude, we have calculated the nuclear Schiff moment in 
nuclei known as relatively soft with respect to the octupole and 
quadrupole excitation modes. We found a considerable enhancement 
of the average magnitude comparable to what was found for nuclei 
having static octupole deformation. Among factors which contribute 
to this enhancement, the main role is played by small energy 
intervals between the opposite parity states and by the large 
vibrational amplitude in soft nuclei. The details of structure of 
the quasiparticle-phonon states are also important. 
 
Our analysis shows that the most favorable conditions can be 
provided by proximity of the quadrupole $2^+$ and octupole $3^-$ 
phonon states, large proton component of the phonon states and 
coupling of the same single-quasiparticle state to quadrupole and 
octupole phonons. We have considered only the lowest symmetric 
quadrupole phonon in the present paper. In the presence of few 
active protons and neutrons, there is a possibility of low-lying 
``mixed-symmetry'' quadrupole phonon in vibrational nuclei that is 
known to be connected to the octupole $3^-$ phonon by a 
considerably stronger E1 strength than the symmetric lowest 
quadrupole phonon \cite{Pietralla}. Therefore one may expect 
additional enhancement if the coupling to the ``mixed-symmetry'' 
phonon is taken into consideration. 
 
The conventional core polarization effects are also important for 
getting a certain quantitative prediction, although they are not 
expected to lead to a significant amplification of the effect. In 
the future the full calculations taking into account on equal 
footing collective and single-particle effects, including the 
${\cal PT}$-violating corrections to the vibrational modes and 
core polarization, are to be carried out. It is desirable to 
extend the studies to lighter nuclei $A \approx 100$ with 
pronounced octupole phonon states with more experimental data 
available. This may help to understand better the nuclear 
structure effects for the enhancement of the Schiff moment and to 
explore a possibility of the correlation between the calculated 
values of the Schiff moment and observed  E1 or E3 strength. 
 
The authors appreciate constructive discussions with N. Auerbach. 
Support from the NSF grants PHY-0070911 and PHY-0244453 is 
gratefully acknowledged. V.F. acknowledges support from the 
Australian Research Council.

\end{document}